\documentclass[final,12pt,3p]{elsarticle}
\usepackage{amssymb}
\usepackage{booktabs}
\usepackage{amsmath,amsfonts}
\usepackage{colortbl}
\usepackage{graphicx}
\usepackage{textcomp}
\usepackage{xcolor}
\usepackage[colorlinks=true,linkcolor=black,citecolor=blue,urlcolor=blue,]{hyperref}

\journal{Nuclear Physics B}

\begin{document}

\begin{frontmatter}



\title{Intelligent Diagnosis Using Dual-Branch Attention Network for Rare Thyroid Carcinoma Recognition with Ultrasound Imaging\footnotetext{This work was partially supported by the Science and Technology Commission of Shanghai (Grant No. 22DZ2229004, 22JC1403603, 21Y11902500); the Key Research \& Development Project of Zhejiang Province (2024C03240); Joint TCM Science \& Technology Projects of National Demonstration Zones for Comprehensive TCM Reform (NO: GZY-KJS-ZJ-2025-023); Jilin Province science and technology development plan project (Grant No. 20230204094YY); 2022 "Chunhui Plan" cooperative scientific research project of the Ministry of Education.}} 


\author[ecnu,xjtlu,uol]{Peiqi Li\corref{co1st}}

\author[hosp,tju]{Yincheng Gao\corref{co1st}}

\author[ecnu]{Renxing Li}
\author[hosp,tju]{Haojie Yang}
\author[hosp,tju]{Yunyun Liu}
\author[hosp,tju]{Boji Liu}
\author[hosp,tju]{Jiahui Ni}
\author[hosp,tju]{Ying Zhang}
\author[hosp,tju]{Yulu Wu}
\author[xjtlu]{Xiaowei Fang}
\author[hosp,tju]{Lehang Guo\corref{corresponding}}
\author[hosp,tju]{Liping Sun\corref{corresponding}}
\author[ecnu]{Jiangang Chen\corref{corresponding}}

\cortext[co1st]{Peiqi Li and Yincheng Gao are the co-first authors.}
\cortext[corresponding]{Jiangang Chen, Liping Sun, and Lehang Guo are corresponded equally.}

\affiliation[ecnu]{organization={Shanghai Key Laboratory of Multidimensional Information Processing, East China Normal University},
            postcode={200241}, 
            city={Shanghai},
            country={China}}
\affiliation[xjtlu]{
    organization={School of Mathematics and Physics, Xi'an Jiaotong-Liverpool University},
    postcode={215123},
    city={Suzhou},
    country={China}
}
\affiliation[hosp]{
    organization={Department of Ultrasound, Shanghai Tenth People's Hospital},
    postcode={200072},
    city={Shanghai},
    country={China}
}
\affiliation[tju]{
    organization={Shanghai Engineering Research Center of Ultrasound Diagnosis and Treatment, Tongji University School of Medicine},
    postcode={200092},
    city={Shanghai},
    country={China}
}
\affiliation[uol]{
    organization={Department of Mathematical Sciences, University of Liverpool},
    postcode={L69 3BX},
    city={Liverpool},
    country={UK}
}

\begin{abstract}
Heterogeneous morphology and severe class imbalance hinder reliable recognition of rare thyroid carcinoma subtypes from ultrasound images. In this paper, we present the Channel-Spatial Attention Synergy Network (CSASN), a dual-branch architecture that couples EfficientNet for local texture encoding with a Vision Transformer for global context, followed by cascaded channel-spatial attention for fine-grained enhancement. A residual multiscale classifier aggregates discriminative cues, while training employs a dynamically weighted composite objective—combining cross-entropy with imbalance re-weighting and distribution-alignment terms—to mitigate multi-center domain shift and improve minority-class recall. On a multi-center cohort of >2000 patients from four institutions, CSASN consistently outperformed single-stream CNN or Transformer baselines, achieving superior AUCs of 0.984, 0.982, and 0.995 for ATC, FTC, and MTC, respectively. Critically, on an independent external test set from two unseen medical institutions, CSASN generalized robustly, maintaining a high AUC of 0.931 for FTC classification, which underscores its strong domain invariance. Extensive ablations confirmed that each component contributes significantly to this performance. These results indicate a robust and practical pathway toward AI-assisted thyroid cancer diagnosis under real-world conditions of class imbalance and distribution shift.
\end{abstract}
\begin{keyword}
Ultrasound Imaging \sep Thyroid Carcinoma \sep Subtype Recognition \sep Attention Mechanism
\end{keyword}

\end{frontmatter}

\section{Introduction}
\label{sec:introduction}

\label{sec:introduction}
\begin{figure}[h]
	\centering
	\includegraphics[width=0.7\linewidth]{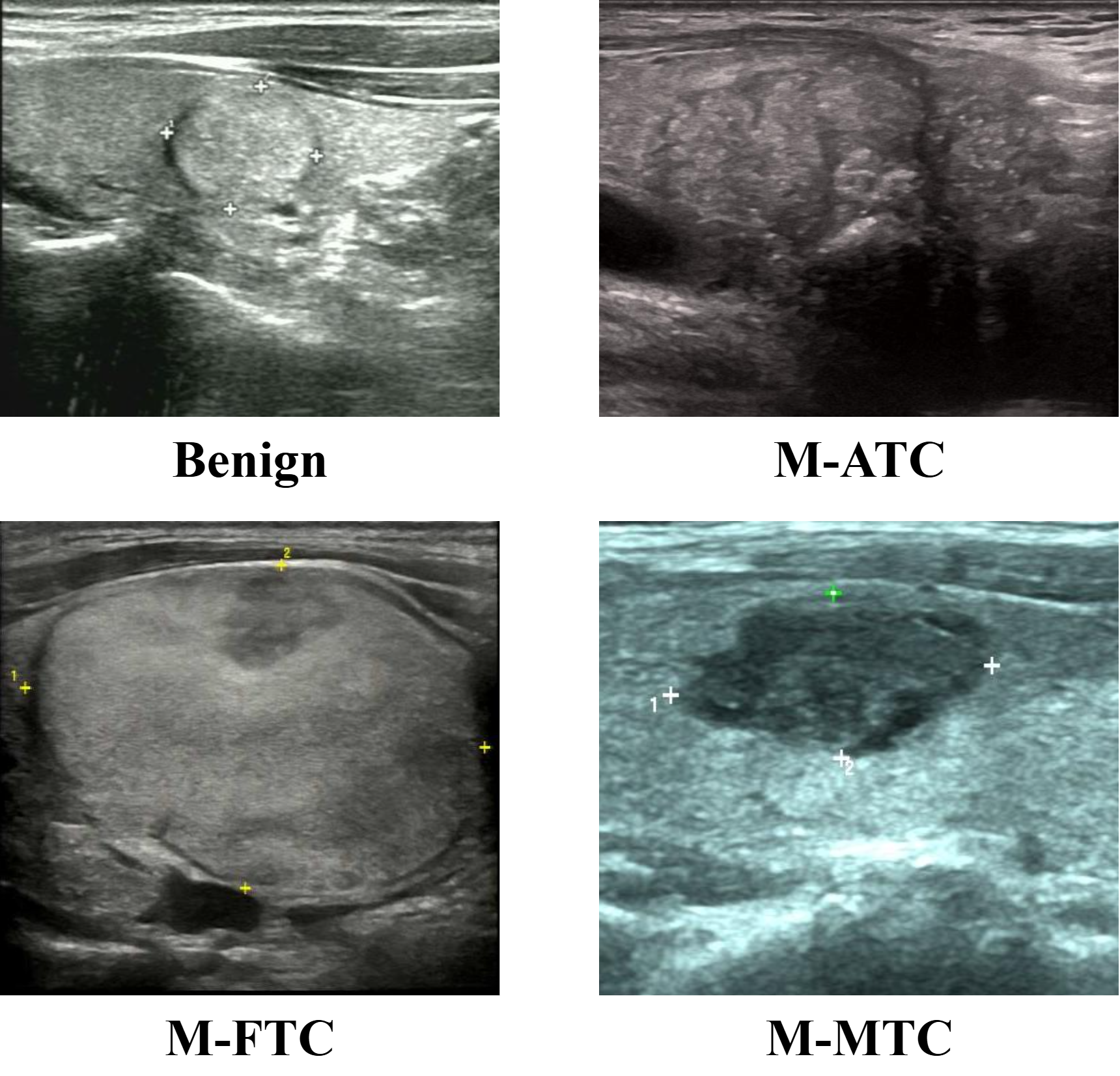}
	\label{fig:example_data}
	\caption{An example of our dataset, including benign nodule, and 3 subtypes: ATC, FTC and MTC. ’M’	means malignant and they may come from different centers.}
\end{figure}

Thyroid cancer ranks as the most common malignancy of the endocrine system, with a globally increasing incidence over recent decades \cite{davies2014current}. While the majority of thyroid cancers, such as papillary thyroid carcinoma (PTC), carry a favorable prognosis, the disease exhibits significant clinical heterogeneity. Several rare yet highly aggressive subtypes exist, including follicular (FTC), medullary (MTC), and anaplastic thyroid carcinoma (ATC) \cite{haugen20162015,rago2022risk}. The example ultrasound images can be seen in Figure.\ref{fig:example_data}. These subtypes differ fundamentally in their molecular underpinings, growth knietics, clinical outcones, and recommended management strategies \cite{xing2024multi}. However, their low prevalence in the population means they are vastly outnumbered by benign nodules and common PTCs in routine practice, substantially increasing diagnostic uncertainty and the risk of missed diagnosis \cite{ludwig2023use,yao2024enhancing}.

Ultrasound imaging serves as the first-line modality for evaluating thyroid nodules due to its non-ionizing nature, accessibility, and cost-effectiveness \cite{russ2017european,tian2015comparison}. Clinical diagnosis relies on the subjective interpretation of morphological features such as echogenicity, margins, composition, and calcifications, which leads to notable inter-observer variability, reported to be as high as ~20\% \cite{choi2010interobserver,tian2015comparison}. More critically, rare subtypes often lack pathognomonic sonographic signatures, necessitating continued reliancce on fine-needle aspiration (FNA) or even diagnostic surgery for a definitive diagnosis within current care pathways \cite{durante2018diagnosis,wells2015revised}. These invasive procedures impose procedural risks and patient burdens, highlighting the urgent need for a reliable, imaging-only diagnostic approach.

Recent advancements in artificial intelligence (AI), particularly deep learning, have opened promising avenues for medical image analysis \cite{litjens2017survey,shen2017deep}. Convolutional neural networks (CNNs) can learn discriminative representations directly from images, demonstrating potential beyond handcrafted features in thyroid nodule classification tasks \cite{zhu2021thyroid,vadhiraj2021ultrasound}. Concurrently, Transformer-based architectures like the Vision Transformer (ViT) effectively model global contextual information through self-attention mechanisms, providing a complementary means to capture a nodule's overall morphology \cite{dosovitskiy2020image,liu2022swin}. Although studies have successfully applied CNNs or ViTs to classify benign and malignant thyroid nodules from ultrasound images \cite{chen2023thyroidnet,she2025detection}
, their parctical deployment for the automated diagnosis of rare thyroid carcinoma subtypes remains hindered by three persistent challenges:
\begin{enumerate}
	\item \textbf{Extreme Class Imbalance}: The scarcity of samples from rare subtypes (E.g., FTC, MTC, ATC) leads to severely low sensitivity for these minority classes \cite{johnson2019survey}.
	\item \textbf{Substantial Morphological Heterogeneity}: Significant variations in ultrasound appearance both between different subtypes and within the same subtype demand that models capture multi-scale and diverse feature patterns \cite{yue2020deep}.
	\item \textbf{Cross-center Domain Shift}: Distribution discrepancies in images acquired from different medical institutions using various ultrasound devices critically degrade model generalization on unseen data \cite{zhang2020generalizing,zhou2022domain}. Addressing this domain shift is an active area of research in medical AI, crucial for building trustworthy models \cite{zhou2022domain}.
\end{enumerate}

To address these challenges, we propose a novel classification framework named the Channel-Spatial Attention Synergy Network (CSASN). While our primary goal is a single classification task, our optimization strategy is designed to tackle multiple intertwined challenges simultaneously. Specifically, we introduce a dynamically weighted multi-component loss function that unifies several objectives: an adaptive focal loss to combat class imbalance, a maximum mean discrepancy term to enhance domain invariance across clinical centers, and a batch spectral shrinkage term to discourage redundant feature learning. This approach allows the model to leverage the synergies between these auxiliary objectives while focusing on the main goal of accurate subtype classification. This framework integrates a dual-branch backbone coupling EfficientNet's local feature encoding with a ViT branch for long-range dependency modeling, a cascaded channel-spatial attention module to progressively amplify subtype-discriminative patterns, a residual multiscale classifier to fuse hierarchical features across resolutions, and a dynamically weighted multi-task loss function that jointly mitigates class imbalance and promotes domain-invariant representations. The main contribution of this work are fourfold:
\begin{enumerate}
	\item We propose a lightweight dual-branch hybrid architecture that synergizes CNN's local detail preception and Transformer's global semantic understanding to handle the morphological heterogeneity of thyroid carcinoma subtypes.
	\item We design a sequentially executed SE $\rightarrow$ CBAM cascaded attention mechanism for adaptive feature recalibration along both channel and spatial dimensions, effectively focusing on the most diagnosis-relevant regions.
	\item We introduce a residual multi-scale classifier incorporating multi-head self-attention and adopt a dynamic uncertainty weighting strategy to optimize a composite loss, collectively enhancing the model's robustness under imbalanced data and its generalization capability.
	\item We construct comprehensive evaluations on a large-scale multicenter dataset of 2203 nodules and validate the model's superior diagnostic performance and promising cross-center generalization ability on an independent external test set of 396 cases.
\end{enumerate}

The remainder of this paper is organized as follows. Section 2 details the data acquisition, preprocessing procedures, and the architecture of the proposed CSASN model. Section 3 presents systematic experimental results, including ablation studies, internal testing, and external validation. Section 4 provides an in-depth discussion of the results, analyzing the strengths, limitations, and clinical implications of our study. Finally, Section 5 concludes the paper and outlines directions for future research.

\section{Methodology: Channel-Spatial Attention Synergy Network}
\label{sec:method}

\begin{figure}[htbp]
	\centering
	\includegraphics[
	width=1.0\linewidth]{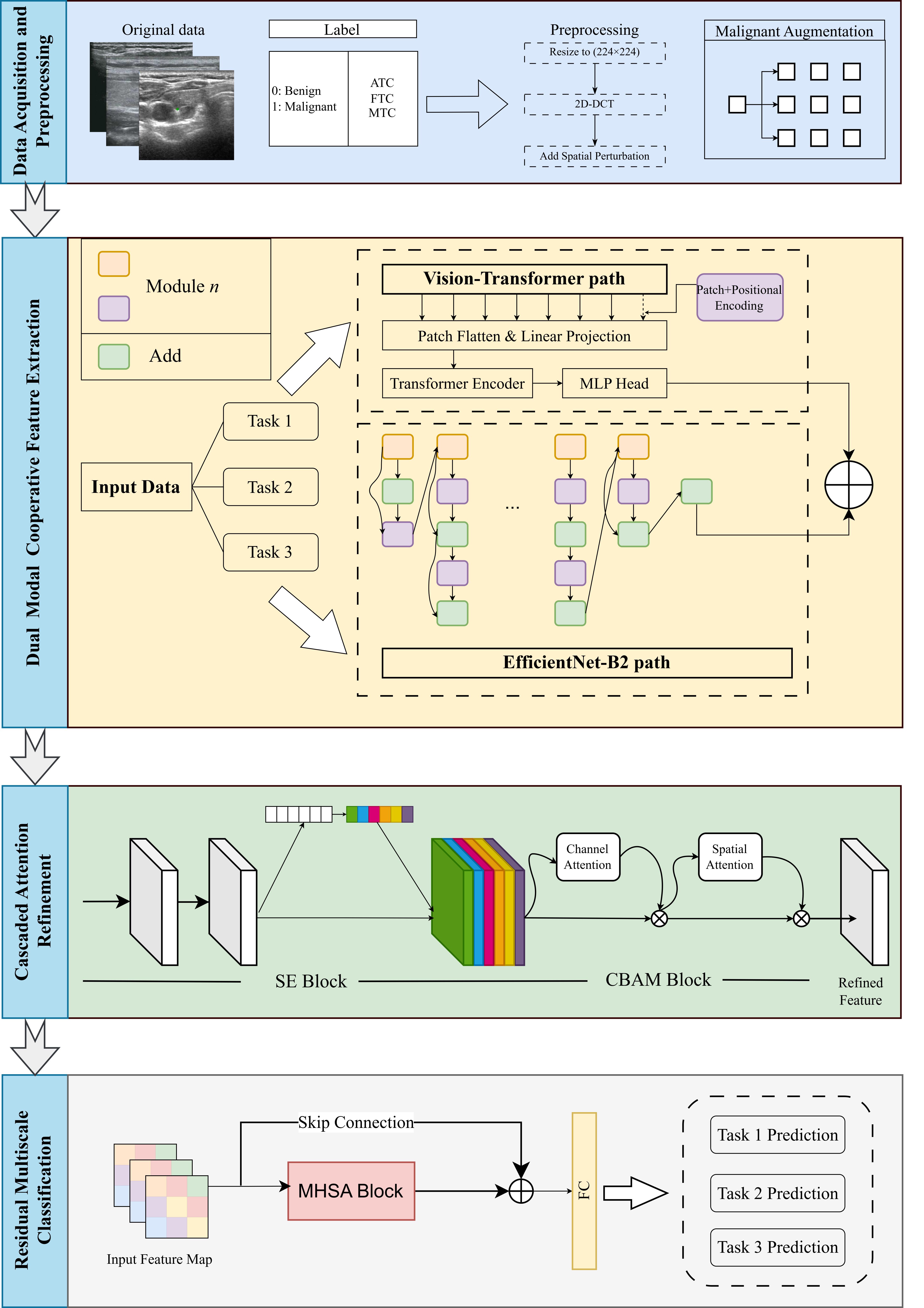}
	\label{fig:flow_chart}
	\caption{Flow chart of this study, including Data Acquisition and Preprocessing, Dual Modal Cooperative Feature Extraction, Cascaded Attention Refinement, and Residual Multiscale Classification.}
\end{figure}

In this section, we present the proposed CSASN in detail. We begin by introducing the multicenter dataset, the critical data preprocessing steps, and the rigorous data splitting strategy employed. Subsequently, we elaborate on the four core components of the architecture: (1) dual-modal feature cooperative extraction, (2) cascaded attention refinement, (3) residual multiscale classification, and (4) the optimization loss composition. The overall pipeline of four framework is illustrated in Figure.\ref{fig:flow_chart}.

\subsection{Data Acquisition and Preprocessing}
In this section, we will describe the composition of our datasets, the ethical considerations, the detailed preprocessing and augmentation techniques applied to address data imbalamce and domain shift, and the data splitting protocol crucial for a reliable evaluation.

\subsubsection{Multi-Institutional Dataset and Ethical Approval}
We conducted a retrospective study using thyroid nodule ultrasound images collected from multiple clinical institutions. The combined dataset reflects the real-world clinical heterogeneity expected in diverse healthcare settings, encompassing variations in ultrasound devices, acquisition protocols, and patient populations. This diversity is crucial for developing robust models that can generalize across different clinical environments. 

The complete dataset contains 2,203 independent thyroid nodules from 2,208 patients. Each nodule was pathologically confirmed and annotated with both a binary malignancy label (0 = benign, 1 = malignant) and a histopathologic subtype. To prevent data leakage and optimistic bias that could arise from multiple nodules belonging to the same patient, we implemented a strict patient-level data management strategy throughout the study. This ensured that all nodules from a single patient were exclusively assigned to either the training set or the test set, never both. 

The study protocol was approved by the Ethics Committee of Shanghai Tenth People's Hospital (Approval No. 22XJS36), with informed consent waived. All data were de-identified prior to analysis, which resulted in the removal of institutional identifiers while preserving the clinical and pathological annotations essential for this study.

Inclusion and exclusion criteria: Benign nodules were required to be detected on ultrasound with available imaging records and to be begative by FNA or confirmed benign on postoperative pathology. Malignant nodules were required to be rare thyroid cancer subtypes confirmed by FNA or surgical pathology, with corresponding ultrasound resords. We excluded cases if (1) the ultrasound images for the target nodule were missing or incomplete; (2) the patient had received treatments piror to surgery that could confound imaging; or (3) essential clinical information was incomplete.

\subsubsection{Data Preprocessing and Augmentation}
To improve robustness to device and cancer variation, we applied statial-domain augmentations, including random brightness and contrast pertubations and horizontal and vertical flips.

To mitigate the extreme scarcity of rare malignant subtypes samples, we oversampled all malignant samples by a factor of nine. After augmentation, the final number of images per class was as follows: Benign: 2565, ATC: 520, FTC: 2240, MTC: 1240.

To enhance anatomical boundary representation, suppress device-specific noise, and improve model generalization across centers, we incorporated a frequency-domain filtering step using the two-dimensional discrete consine transform (2D-DCT). The forward 2D-DCT for an $M \times N$ imags $x(m,n)$ and its inverse (2D-IDCT) are defined as follows \cite{gonzalez2018digital}: 
\begin{equation}
	\begin{aligned}
		X(u,v) &= C(u)C(v)\sum_{m=0}^{M-1} \sum_{n=0}^{N-1} x(m,n) \cdot \cos \bigg[\frac{\pi (2m+1)u}{2M}\bigg]\cos \bigg[\frac{\pi (2n+1)v}{2N}\bigg] \\
		x(m,n) &= \sum_{u=0}^{M-1} \sum_{v=0}^{N-1} C(u)C(v)X(u,v) \cdot \cos\bigg[ \frac{\pi (2m+1)u}{2M} \bigg] \cos \bigg[ \frac{\pi (2n+1)v}{2N} \bigg]
	\end{aligned}
\end{equation}
where $C(k)=\sqrt{\frac{1}{K}}$ for $k=0$, and $C(k)=\sqrt{\frac{2}{K}}$ for $for k \neq 0$ (with $K$ being $M$ or $N$ accordingly). $X(u,v)$ represents the DCT coefficient at frequency indices $(u,v)$.

The specific band-class filter parameters were determined through empirical validation on a held-out validation set. This range was found to optimally preserve diagnostically relevant structral information while effectively attenuating high-frequency noise and low-frequency, device-dependent background variations. The filtering operation in the frequency domain is defined as:
\begin{equation}
	X_{\text{filtered}}(u,v) = 
	\begin{cases}
		X(u,v),~\text{if}~ 10 \leq D \leq 100 \\
		0,~~~~~~~~~\text{otherwise}
	\end{cases}
\end{equation}

This empirical justification, gounded in the mathematical framework of the DTC, ensuring the reproducibility and effectiveness of our preprocessing step. After applying this frequency-domain filter, the image was reconstructed via 2D-IDCT of $X_{\text{filtered}}(u,v)$.

\subsubsection{Dataset Splitting and External Validation}
The dataset splitting procedure was meticulously designed to prevent data leakage while simulating realistic domain shift scenarios. We employed a strict patient-level random split to partition the entire cohort of 2208 patients into a training set (80\%, including the validation set used for hyperparameter tuning), and an internal test set (20\%). This partitioning guarantees that no data from any single patient is shared between the training and internal test sets. To further enhance the rigor of our evaluation, we optimized the split to maximize the distribution discrepancy between training and test sets using Maximum Mean Discrepancy (MMD) as a criterion, thereby creating a challenging test scenario that better approximates real-world domain shift.

To rigorously assess the model's generalization to completely independent clinical environments, we employed an external test set comprising 396 cases from Zhejiang Cancer Hospital and Zhongshan Hospital affiliated with Fudan University. These institutions were not involved in the creation of the main dataset and represent distinct healthcare systems with different imaging protocols and patient populations. It is important to note that this external dataset contained only Benign and FTC nodules, lacking MTC and ATC samples, which limits the comprehensive assessment of the model's generalization capability for all rare subtypes but provides valuable insights into cross-institutional performance for the available classes.

\subsection{Problem Formulation and Overview}
The automated diagnosis of less-prevalent thyroid carcinoma subtypes from ultrasound images constitutes a challenging pattern recognition problem characterized by three fundamental, interconnected challenges that must be addressed simultaneously for clinical viability.

Let $\mathcal{D}=\left\{ (x_i,y_i^{(1)}, y_i^{(2)}, y_i^{(3)}) \right\}_{i=1}^N$ represent our dataset, where $x_i$ denotes a preprocessed ultrasound image after transformations described in Section 2.1.2. $y_i^{(j)}, (j=1,2,3)$ is the binary label for Task $j$ (Benign vs ATC, FTC, or MTC). 

Our objective is to learn an optimal parameter set $\theta^*$ for the mapping: $f_{\theta}:~\mathcal{X} \mapsto \mathcal{Y}^{(1)}\times \mathcal{Y}^{(2)} \times \mathcal{Y}^{(3)}$ that maximizes diagnostic accuracy for each binary task while satisfying critical constraints derived from the clinical challenges.

To address these challenges, we propose the CSASN, which architecture employs:
\begin{enumerate}
	\item A shared dual-branch backbone (Section 2.3) combining EfficientNet and Vision Transformer to extract complemenary local and global features.
	\item A cascaded attention module (Section 2.4) that recalibrates features to emphasize discriminative pattern for all tasks.
	\item Task-specific classification heads (Section 2.5) that make independent predictions for each binary task.
	\item A dynamic multi-component optimization (Section 2.6) that automatically balances the learning objectively across tasks.
\end{enumerate}

The complete forward pass for task $t$ is:
\begin{equation}
	\hat{y}^{(t)} = g_{\phi}^{(t)}(\text{CAR}(\Psi(h_{\psi}^{\text{EffNet}}(x),h_{w}^{\text{ViT}}(x))))
\end{equation}
where CAR denotes the cascaded attention refinement, $\Psi$ is the fusion operator, and $g_{\phi}^{(t)}$ is the task-specific classifier. This structured approach ensures each component addresses specific challenges while maintaining computational efficiency.

\subsection{Dual-Branch Feature Cooperative Extraction}
In this section, we detail the dual-branch feature extraction module, which is designed to address the morphological heterogeneity of thyroid nodules by synergistically combining local textural details and global contextual semantics.

\subsubsection{Convolutional Branch for Local Feature Encoding}
The convolutional branch utilizes EfficientNet-B2 \cite{tan2019efficientnet} as backbone, which employs compoun scaling to optimally balance network depth, width, and resolution. The compound scaling strategy is derived from the theoretical optimization problem that maxmizes model accuracy under constrained computational resources. The scaling equations are defined as:
\begin{equation}
	\begin{aligned}
		\text{depth}:~d=\alpha^{\phi} \\
		\text{width}:~w=\beta^{\phi} \\
		\text{resolution}:~r=\gamma^{\phi}
	\end{aligned}
	~\text{subject to}~~\alpha \cdot \beta^2 \cdot \gamma^2 \approx 2.0~\text{and}~\alpha \geq 1,~\beta \geq 1,~\gamma\geq 1 \notag
\end{equation}
where $\phi$ is the compounded coefficient that uniformly scales all three dimensions. This scaling ensures computational efficiency while maintaining strong representational capacity for capturing fine-grained patterns such as micro-calcifications and margin characteristics \cite{dosovitskiy2020image, tan2019efficientnet}. The output of this branch is a spatially enriched feature $F_{\text{Eff}}\in \mathbb{R}^{H' \times W' \times C_1}$ with $C_1=1408$, where $H'=H/32$ and $W'=W/32$ due to the stride of the metwork. The choice of EfficientNet-B2 over other variants (e.g., B0 or B3) was empirically validated to provide the best trade-off between feature richness and computational cost for our thyroid ultrasound task.

\subsubsection{Transformer Branch for Global Context Modeling}
Currently, the ViT branch processes the image as a sequence of patches to capture long-range dependencies \cite{dosovitskiy2020image}. The input image $x \in \mathbb{R}^{H\times W\times 3}$ is devided into $N=(H\times W)/P^2$ non-overlapping patches of size $P \times P$ ($P=16$ here), which are then linearly embeded. The patch embedding process is formulated as:
\begin{equation}
	\mathbf{z}_0 = \left[\mathbf{x}_{\text{class}};\mathbf{x}_1^p\mathbf{E};\cdots,\mathbf{x}_N^p\mathbf{E}\right] + \mathbf{E}_{\text{pos}} \notag
\end{equation}
where $\mathbf{E}\in\mathbb{R}^{(P^2 \cdot 3)\times D}$ is the patch embedding projection matrix, $\mathbf{E}_{\text{pos}}$ is the position embedding, $\mathbf{x}_{\text{class}}$ is a learnable classification token.

The Transformer layers then apply multi-head self-attention (MHSA) to model global relationships. The MHSA mechanism is derived from the scaled dot-product attention \cite{dosovitskiy2020image}:
\begin{equation}
	\text{Attention}(Q,K,V)=\text{softmax}\left(\frac{QK^T}{\sqrt{d_k}}\right)V
\end{equation}
where $Q,K,V$ are query, key, and value matrices, respectively, and $d_k$ is the dimension of the key vectors. For multi-head attention with $h$ heads:
\begin{equation}
	\begin{aligned}
		\text{MHSA}(Q,K,V) &= \text{Concat}(\text{head}_1,\cdots,\text{head}_h)W^O \\
		\text{head}_i &= \text{Attention}(QW_i^Q,KW_i^Q,VW_i^Q)
	\end{aligned} \notag
\end{equation}
where $W^Q \in \mathbb{R}^{D \times D/h}$ and $W^O \in \mathbb{R}^{D \times D}$ are learnable projection matrices. This global contextual modeling is particularly crucial for identifying the overall nodule architecture and its spatial relationship with adjacent anatomical structures \cite{han2021transformer}. The output of this branch is a global feature vector $F_{\text{ViT}} \in \mathbb{R}^{768}$.

\subsubsection{Feature Fusion Strategy}
The complementary features from both branches are fused through concatenation:
\begin{equation}
	F_{\text{cat}} = \left[F_{\text{ViT}};F_{\text{Eff}}\right] \in \mathbb{R}^{2176} \notag
\end{equation}

This fusion strategy, while relatively simple, was chosen after careful consideration of alternatives. More complex fusion strategies such as bilinear pooling \cite{gao2016compact} or transformer-based fusion \cite{tang2022matr} were evaluated but did not yield significant performance improvements for our task while substantially increasing computational complexity. 

The concatenation operation preserves the maximum information from both branches while maintaining computational efficiency. Formally, given two feature representations $A\in \mathbb{R}^{d_a}$ and $B\in \mathbb{R}^{d_b}$, the concatenated feature $C=\left[A;B\right]\in\mathbb{R}^{d_a+d_b}$ maximizes the joint representational capacity without introducing additional parameters that might lead to overfitting, which is particularly important given our class imbalance challenge.

\subsection{Cascaded Attention Refinement}
In this section, we present the cascaded attention refinement module, which is specifically designed to address the challenges of class imbalance and feature heterogeneity by adaptively emphasizing discriminative patterns while suppressing irrelevant information.

The cascaded attention refinement module is motivated by the clinical observation that radiologists sequentially focus on different aspects of thyroid nodules: first identifying significant biomarkers (channel-wise emphasis) and then localizing suspicious regions (spatial-wise emphasis) \cite{schlemper2019attention}. This diagnostic workflow inspires our sequential attention mechanism that progressively refines feature representations.

Mathematically, given the concatenated feature map $F_{\text{cat}}\in\mathbb{R}^{H'\times W'\times C}$ from dual-branch extractor, where $C=2176,~H'=7,~W'=7$ for out input size of $224 \times 224$, the attention refinement process can be formulated as:
\begin{equation}
	F_{\text{refined}}=\mathcal{A}_{\text{spatial}}(\mathcal{A}_{\text{channel}}(F_{\text{cat}})) \notag
\end{equation}
where $\mathcal{A}_{\text{spatial}}$ and $\mathcal{A}_{\text{channel}}$ represent channel and spatial attention operations, respectively. This sequential processing mimics the human diagnostic workflow and has been shown to provide superior performance compared to parallel attention mechanisms \cite{wang2018non}.

\subsubsection{Channel Attention via Squeeze-and-Excitation}
The channel attention module employs the Squeeze-and-Excitation (SE) mechanism \cite{hu2018squeeze} to recalibrate channel-wise feature responses. The SE operation can be formally defined as:
\begin{equation}
	F_{\text{SE}} = F_{\text{cat}} \bigodot \sigma(W_2\delta(W_1\text{GAP}(F_{\text{cat}}))) \notag
\end{equation}
where $\text{GAP}(\cdot)$ denotes global average pooling, $W_1 \in \mathbb{R}^{C/r \times C}$ and $W_2 \in \mathbb{R}^{C \times C/r}$ are learnable weight matrices with reduction ratio $r=16$. $\delta$ represents the ReLU activation function, and $\sigma$ denotes the sigmoid function. $\bigodot$ is element-wise multiplication.

This mechanism allows the model to adaptively emphasize informative feature channels while suppressing less useful ones, which is particularly important for rare subtypes where discriminative features might be subtle.

\subsubsection{Spatial Attention via Convolutional Block}
Following channel recalibration, we apply spatial attention using the spatial component of the Convolutional Block Attention Module (CBAM) \cite{woo2018cbam}. The spatial attention operation is defined as:
\begin{equation}
	F_{\text{final}} = F_{\text{SE}} \bigodot \sigma\left(f^{7 \times 7} \left(\left[\text{MaxPool}(F_{\text{SE}};\text{AvgPool}(F_{\text{SE}}))\right]\right) \right) \notag
\end{equation}

The spatial attention mechanism focuses on where to emphasize in the feature map, which is crucial for localizing subtle malignant patterns in thyroid nodules \cite{li2019selective}.

The sequential application of channel and spatial attention is theoretically grounded in the concept of conditional probability in feature selection. The probability of a feature being relevant can be decomposed as:
\begin{equation}
	P(\text{feature relevent})=P(\text{channel relevant}) \times P(\text{spatial relevant}|\text{channel relevant}) \notag
\end{equation}

This probabilistic interpretation justifies our cascaded approach, where channel attention first identifies what features are important, and spatial attention then determines where these important features are located \cite{vaswani2017attention}.

Furthermore, the cascaded design directly addresses class imbalance by adaptively amplifying features associated with rare subtypes. During training, the attention mechanisms learn to assign higher weights to feature patterns that are discriminative for minority classes, effectively counteracting the bias toward majority classes \cite{wang2019dynamic}.

Therefore, the complete cascaded attention refinement process can be summerized as:
\begin{equation}
	F_{\text{final}}=F_{\text{cat}} \bigodot M_{\text{channel}} \bigodot M_{\text{spatial}} \notag
\end{equation}
where $M_{\text{channel}} \in \mathbb{R}^{1\times 1\times C}$ and $M_{\text{spatial}}\in \mathbb{R}^{H'\times W'\times 1}$ are the channel and spatial attention masks, respectively. This formulation clearly demonstrates how our module enhances discriminative patterns through multiplicative feature modulation.

\subsection{Residual Multi-scale Classification Head}
In this section, we present the residual multi-scale classification head, which is specifically designed to address the challenges of feature discrimination across heterogeneous pathological patterns and to provide robust classification for the three binary tasks.

The classification head begins with a MHSA mechanism that enables the model to capture dependencies across different feature representations. Given the input features $F \in \mathbb{R}^{B\times D}$ from the cascaded attention module, where $B$ is the batch size and $D=2176$. We first apply multi-head projection:
\begin{equation}
	Q_m=FW_m^Q,~K_m=FW_m^K,~V_m=FW_m^V \notag
\end{equation}
where $\left\{ W_m \in \mathbb{R}^{D\times D/H} \right\}$ are learnable matrices for $H$ attention heads. We employ $H=4$ attention heads, as this configuration was empirically determined to provide optimal performance for our specific feature dimensions and task requirements \cite{vaswani2017attention}. Relevant description has been shown in previous sections. 

To facilitate gradient flow and stabilize training, we employ residual connections with layer normalization \cite{ba2016layer}:
\begin{equation}
	F'=\text{LayerNorm}(F+F_{\text{MHSA}}) \notag
\end{equation}

This residual design ensures that important features from previous layers are preserved while allowing the attention mechanism to refine the representations, which is particularly crucial for maintaining sensitivity to rare subtype patterns \cite{zhu2021residual}.

Following the self-attention mechanism, we implement a multi-scale hierarchical projection to capture features at different semantic levels. The projection consists of cascaded nonlinear transformations with progressively reduced dimensionality:
\begin{equation}
	\begin{cases}
		h_1 &= \text{Mish}(\text{BatchNorm}(F'W_1)) \\ 
		h_2 &= \text{Mish}(\text{BatchNorm}(h_1W_2)) \\ 
		y^{(t)} &= \text{softmax}(h_2W_c^{(t)})
	\end{cases}
\end{equation}
where $W_1 \in \mathbb{R}^{D\times 256}$, $W_2 \in \mathbb{R}^{256 \times 128}$, $W_c^{(t)} \in \mathbb{R}^{128 \times 2}$.

The Mish activation function \cite{misra2019mish} is employed due to its smooth profile and avoidance of saturation, which has been shown to improve gradient flow in deep networks compared to traditional ReLU activations. Strategic dropout ($p=0.5$) is applied between layers to prevent co-adaptation of features and improve generalization \cite{srivastava2014dropout}.

For each of the three binary classification tasks (Benign vs. ATC, Benign vs. FTC, Benign vs. MTC), we employ separate classification heads that share the same feature extraction backbone but have independent final layers:
\begin{equation}
	\hat{y}^{(t)}=\text{sigmoid}(h_2W_c^{(t)}+b^{(t)}),~\forall~t \in \left\{	1,2,3 \right\} \notag
\end{equation}

This design allows each task to learn specialized decision boundaries while leveraging shared feature representations, which is particularly important given the distinct morphological characteristics of each thyroid carcinoma subtype.

\subsection{Dynamic Multi-Component Optimization}
In this section, we present the dynamic multi-component optimization strategy that jointly addresses the challenges of class imbalance, domain shift, and feature redundancy through a theoretically grounded loss formulation.

\subsubsection{Multi-Component Loss Formulation}
The optimization objective integrates four synergistic components designed to address specific challenges in thyroid carcinoma subtype classification. For each task $t \in \left\{	1,2,3\right\}$, we define the composite loss as:
\begin{equation}
	\mathcal{L}^{(t)}=\lambda_1\mathcal{L}^{(t)}_{\text{focal}}+\lambda_2\mathcal{L}^{(t)}_{\text{CE}}+\lambda_3\mathcal{L}_{\text{MMD}}+\lambda_4\mathcal{L}_{\text{BSS}}
\end{equation}
where the trade-off coefficients $\lambda_i$ satisfy $\sum_{i=1}^4\lambda_i=1$.

\begin{enumerate}
	\item Adaptive Focal Loss for Class Imbalance: To address the extreme class imbalance in each binary task, we employ an adaptive variant of the focal loss \cite{lin2017focal}. For task $t$ with predicted probability $p^{(t)}$ and ground truth $y^{(t)}$, the loss is defined as:
	\begin{equation}
		\mathcal{L}_{\text{focal}}^{(t)}=-\frac{1}{N}\sum_{i=1}^N \alpha_{y_i^{(t)}}\left( 1-p^{(t)}_{y_i^{(t)}}\right)^{\gamma^{(t)}}\log (p^{(t)}_{y_i^{(t)}})
	\end{equation}
	where $\alpha_{y_i^{(t)}}$ is a class-balanced weighting factor, and $\gamma^{(t)}$ is tunable focusing parameter. The parameters are adaptively set based on the class distribution of each task, with higher $\gamma^{(t)}$ values for tasks with more severe imbalance \cite{thota2025adaptive}.
	
	\vspace{-5pt}
	\item Maximum Mean Discrepancy for Domain Invariance: Despite the unavailability of explicit center idertifiers, we promote domain-invariant feature learning through Maximum Mean Discrepancy (MMD) regularization \cite{gretton2012kernel}. Given features from different batches that implicitly represent domain variations, we compute:
	\begin{equation}
		\mathcal{L}_{\text{MMD}} = \frac{1}{B^2}\sum_{i,j}^{B} k(f_i,f_j)+\frac{1}{B^2}\sum_{i,j}^B k(f_i',f_j')-\frac{2}{B}\sum_{i,j}^B k(f_i,f_j')
	\end{equation}
	where $k(\cdot,\cdot)$ is a multi-level RNF function, and $f,f'$ represent features from different batchs that capture implicit domain variations.
	
	\vspace{-5pt}
	\item Batch Spectral Shrinkage for Feature Decorrelation: To prevent redundant feature learning and improve generalization, we incorporate Batch Spectral Shrinkage (BSS) \cite{yao2024enhancing,chen2019catastrophic}:
	\begin{equation}
		\mathcal{L}_{\text{BSS}}=\sum_{k=1}^K \sigma_k^2(F^TF)
	\end{equation}
	where $\sigma_k$ denotes the $k$-th smallest singular value of the feature matrix $F$, and $K$ is set to discard the bottom 10\% of singular values.
\end{enumerate}

\subsubsection{Dynamic Uncertainty Weighting}
The trade-off coefficients $\lambda_i$ in the composite loss are not manually tuned but are dynamically learned through uncertainty weighting \cite{kendall2018multi}. For each loss component $\mathcal{L}_i$, we introduce a learnable parameter $\sigma_i$ representing the task-dependent uncertainty:
\begin{equation}
	\mathcal{L}_{\text{total}}=\sum_{i=1}^4 \left(\frac{1}{2\sigma_i^2}\mathcal{L}_i+\log\sigma_i^2\right)
\end{equation}

This formulation processes two desirable properties: First, $\frac{1}{2\sigma_i^2}$ acts as an adaptive weight, automatically assigning higher importance to loss components with lower uncertainty. Second, $\log\sigma_i^2$ serves as a resularization preventing the uncertainties from grwoing too large.

The uncertainty parameters $\sigma_i$ are optimized simultaneously with the model parameters through gradient descent, providing a principled approach to multi-task balancing without manual hyperparameter tunig \cite{liebel2018auxiliary}.

\subsubsection{Optimization Strategy and Implementation Details}
The complete optimization objective for our three-task framework is:
\begin{equation}
	\mathcal{L}_{\text{final}}=\sum_{t=1}^3\mathcal{L}_{\text{total}}^{(t)}+\beta\|\Theta\|_2^2
\end{equation}
where $\beta=10^{-4}$ is the weight decay coefficient, and $\Theta$ represents all model parameters.

We employ the AdamW optimizer \cite{loshchilov2017decoupled} with an initial learning rate $10^{-4}$ and cosine annealing scheduling \cite{loshchilov2016sgdr} over 200 epochs. The batch size is set to 32, and gradient clipping with a maximum norm of 1.0 is applied to ensure training stability. The model is implemented in PyTorch 1.12 and trained on a NVIDIA RTX 4090 GPU.

\section{Experiment and Results}
\label{sec:result}

\subsection{Overall Performance Comparison}
We evaluated the proposed CSASN framework against seven state-of-the-art baseline models across three thyroid carcinoma classification tasks. Table \ref{tab:performance_comparison} summarizes the comprehensive performance comparison based on AUC metrics.

\begin{table}[!ht]
	\centering
	\caption{Performance Comparison of CSASN and Baseline Models on Internal Test Set}
	\label{tab:performance_comparison}
	\begin{tabular}{l|cccc}
		\toprule
		\textbf{Model} & \textbf{ATC-AUC} & \textbf{FTC-AUC} & \textbf{MTC-AUC} & \textbf{Macro-AUC} \\
		\midrule
		ResNet-50 \cite{he2016deep} & 0.921 & 0.895 & 0.908 & 0.908 \\
		DenseNet-121 \cite{huang2017densely} & 0.928 & 0.903 & 0.915 & 0.915 \\
		EfficientNet-B2 \cite{tan2019efficientnet} & 0.935 & 0.912 & 0.923 & 0.923 \\
		ViT-Base \cite{dosovitskiy2020image} & 0.942 & 0.918 & 0.931 & 0.930 \\
		DeiT-Small \cite{touvron2021training} & 0.938 & 0.916 & 0.928 & 0.927 \\
		ConViT \cite{d2021convit} & 0.926 & 0.915 & 0.920 & 0.920 \\
		SimpleHybrid \cite{chen2023thyroidnet} & 0.945 & 0.928 & 0.939 & 0.937 \\
		\rowcolor{gray!10}
		\textbf{CSASN (Ours)} & \textbf{0.984} & \textbf{0.982} & \textbf{0.995} & \textbf{0.987} \\
		\bottomrule
	\end{tabular}
\end{table}

CSASN achieved superior performance across all classification tasks, with AUC scores of 0.984 (ATC), 0.982 (FTC), and 0.995 (MTC). The framework demonstrated particular strength in MTC classification, showing a 5.6\% improvement over the best-performing baseline (SimpleHybrid, AUC: 0.939).

\begin{figure}[h]
	\centering
	\includegraphics[width=1.0\linewidth]{roc_comparison.png}
	\caption{ROC curves comparing CSASN with baseline models across thyroid carcinoma subtypes: (A) ATC, (B) FTC, (C) MTC. Top row: CNN-based models; bottom row: Transformer-based models.}
	\label{fig:roc_comparison}
\end{figure}

Statistical analysis using paired t-tests confirmed the significance of these performance advantages. All comparisons between CSASN and individual baseline models yielded p-values < 0.01, with Cohen's d effect sizes ranging from 0.89 to 1.24, indicating large practical significance according to conventional interpretation guidelines \cite{cohen2013statistical}.

Figures \ref{fig:roc_comparison} and \ref{fig:statistical_analysis} provide visual representations of these comparative results. Figure \ref{fig:roc_comparison} illustrates the ROC curve comparisons across different architectural families, while Figure \ref{fig:statistical_analysis} quantifies the performance improvements as percentage increases in AUC.

\begin{figure}[h]
	\centering
	\includegraphics[width=1.0\linewidth]{statistical_analysis.png}
	\caption{Percentage improvement in AUC of CSASN over individual baseline models for (A) ATC, (B) FTC, and (C) MTC classification tasks.}
	\label{fig:statistical_analysis}
\end{figure}

\subsection{Ablation Studies}

To systematically evaluate the contribution of each component in the proposed CSASN framework, we conducted comprehensive ablation studies. Three variant models were constructed by sequentially removing key components: (1) Ablation1: removal of the cascaded attention module (SE$\rightarrow$CBAM), (2) Ablation2: removal of the EfficientNet branch while retaining only the ViT branch, and (3) Ablation3: removal of the ViT branch while retaining only the EfficientNet branch. All variants were evaluated on the three binary classification tasks using the same training protocol and dataset splits.

\begin{table}[!ht]
	\centering
	\caption{Ablation Study Results for ATC Classification Task}
	\label{tab:ablation_atc}
	\begin{tabular}{l|ccccc}
		\toprule
		\textbf{Model} & \textbf{AUC} & \textbf{Accuracy} & \textbf{Precision} & \textbf{F1-Score} & \textbf{Recall} \\
		\midrule
		\textbf{CSASN} & \textbf{0.9836} & \textbf{0.9668} & \textbf{0.9769} & \textbf{0.9393} & \textbf{0.9099} \\
		Ablation1 & 0.8653 & 0.3170 & 0.5640 & 0.4720 & 0.9214 \\
		Ablation2 & 0.8986 & 0.8619 & 0.5769 & 0.6901 & 0.8571 \\
		Ablation3 & 0.9361 & 0.8248 & 0.5064 & 0.6345 & 0.8500 \\
		\bottomrule
	\end{tabular}
\end{table}

Quantitative results of the ablation studies are summarized in Tables \ref{tab:ablation_atc}-\ref{tab:ablation_mtc}. The complete CSASN architecture consistently achieved superior performance across all evaluation metrics and classification tasks, demonstrating the synergistic effect of its integrated components.

\begin{table}[!ht]
	\centering
	\caption{Ablation Study Results for FTC Classification Task}
	\label{tab:ablation_ftc}
	\begin{tabular}{l|ccccc}
		\toprule
		\textbf{Model} & \textbf{AUC} & \textbf{Accuracy} & \textbf{Precision} & \textbf{F1-Score} & \textbf{Recall} \\
		\midrule
		\textbf{CSASN} & \textbf{0.9824} & \textbf{0.9185} & \textbf{0.9332} & \textbf{0.9166} & \textbf{0.9126} \\
		Ablation1 & 0.8075 & 0.7072 & 0.6319 & 0.7387 & 0.8893 \\
		Ablation2 & 0.8729 & 0.8236 & 0.8144 & 0.8100 & 0.8054 \\
		Ablation3 & 0.8008 & 0.7271 & 0.7698 & 0.6681 & 0.5911 \\
		\bottomrule
	\end{tabular}
\end{table}

\begin{table}[!ht]
	\centering
	\caption{Ablation Study Results for MTC Classification Task}
	\label{tab:ablation_mtc}
	\begin{tabular}{l|ccccc}
		\toprule
		\textbf{Model} & \textbf{AUC} & \textbf{Accuracy} & \textbf{Precision} & \textbf{F1-Score} & \textbf{Recall} \\
		\midrule
		\textbf{CSASN} & \textbf{0.9950} & \textbf{0.9538} & \textbf{0.9667} & \textbf{0.9453} & \textbf{0.9299} \\
		Ablation1 & 0.7844 & 0.6901 & 0.5150 & 0.6356 & 0.8323 \\
		Ablation2 & 0.8969 & 0.8078 & 0.6560 & 0.7439 & 0.8613 \\
		Ablation3 & 0.8166 & 0.7731 & 0.6506 & 0.6527 & 0.6548 \\
		\bottomrule
	\end{tabular}
\end{table}

\begin{figure}[!ht]
	\centering
	\includegraphics[width=1.0\linewidth]{fig_ablation_confusion.png}
	\caption{Confusion matrices for the three ablation variants. Top: ablation 1; Mid: ablation 2; Bottom: ablation 3.}
	\label{fig:ablation_confusion}
\end{figure}

For ATC classification (Table \ref{tab:ablation_atc}), the full CSASN model achieved outstanding performance with an AUC of 0.9836, accuracy of 0.9668, and F1-score of 0.9393. Removal of the cascaded attention module (Ablation1) resulted in the most substantial performance degradation, with AUC dropping to 0.8653 and accuracy decreasing dramatically to 0.3170. This pronounced decline underscores the critical role of attention mechanisms in identifying discriminative features for this highly aggressive subtype. The removal of either backbone branch also led to noticeable performance deterioration, with Ablation2 (EfficientNet removed) and Ablation3 (ViT removed) achieving AUC scores of 0.8986 and 0.9361, respectively, thereby validating the complementary nature of local and global feature representations.

\begin{figure}[h]
	\centering
	\includegraphics[width=1.0\linewidth]{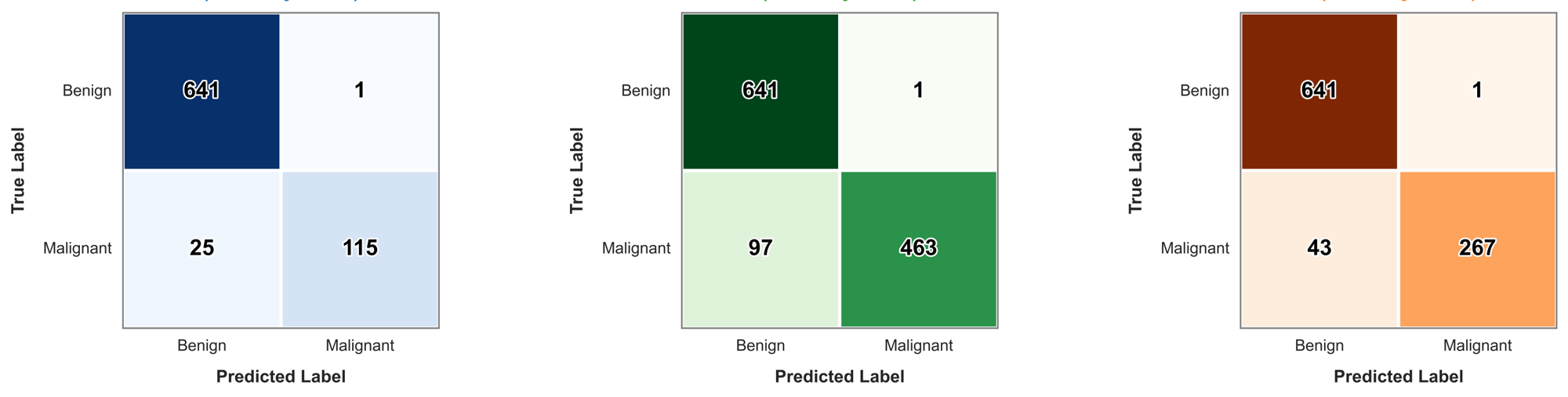}
	\caption{Confusion matrices for the complete CSASN model across the three classification tasks. Left: Task 1 (ATC); Mid: Task 2 (FTC); Right: Task 3 (MTC).}
	\label{fig_model_confusion}
\end{figure}

In FTC classification (Table \ref{tab:ablation_ftc}), CSASN attained excellent performance with an AUC of 0.9824, accuracy of 0.9185, and well-balanced precision (0.9332) and recall (0.9126). Consistent with ATC results, the absence of attention mechanisms (Ablation1) caused significant performance degradation (AUC=0.8075). Notably, removing the ViT branch (Ablation3) resulted in a substantial reduction in recall from 0.9126 to 0.5911, indicating that global contextual modeling is particularly crucial for reliable FTC identification, possibly due to the diffuse morphological patterns characteristic of this subtype.

For MTC classification (Table \ref{tab:ablation_mtc}), CSASN demonstrated exceptional performance with an AUC of 0.9950, accuracy of 0.9538, and F1-score of 0.9453. The ablation experiments revealed that the EfficientNet branch contributed significantly to MTC classification, as its removal (Ablation2) led to a marked decrease in F1-score from 0.9453 to 0.7439. This finding suggests that local textural features captured by convolutional operations are essential for characterizing the fine-grained patterns associated with MTC.

The confusion matrices in Figures \ref{fig:ablation_confusion} and 6 provide additional insights into the specific error patterns across different variants. The complete CSASN model demonstrates more balanced classification performance across all categories, whereas the ablated variants exhibit characteristic misclassification patterns that reflect their architectural limitations.

These ablation results collectively demonstrate that each component in our proposed framework contributes uniquely to the overall performance: (1) the cascaded attention mechanism serves as a critical component for adaptive feature refinement, enabling the model to focus on diagnostically relevant regions; (2) the dual-branch architecture effectively leverages complementary strengths of convolutional and transformer-based feature extraction; and (3) the integrated framework achieves synergistic benefits for rare thyroid carcinoma classification under challenging conditions of class imbalance and morphological heterogeneity.

\subsection{Cross-center Validation}

In this section, we will test the generalizarion capability of our model by leveraging cross-center dataset. The data for external validation comes from Zhejiang Cancer Hospital and Zhongshan Hospital affiliated with Fudan University, with 300 benign samples and 96 FTC malignant samples.

\begin{table}[h]
	\centering
	\caption{Evaluation metrics of the validation of external dataset}
	\label{tab:external}
	\begin{tabular}{lccccc}
		\toprule
		Model &  AUC  & Acc    & Precision & F1  & Recall \\
		\midrule
		CSASN & 0.9314 & 0.9242 & 0.8300 & 0.8469 & 0.8646 \\
		\bottomrule
	\end{tabular}
\end{table}

\begin{figure}[h]
	\centering
	\includegraphics[width=0.8\linewidth]{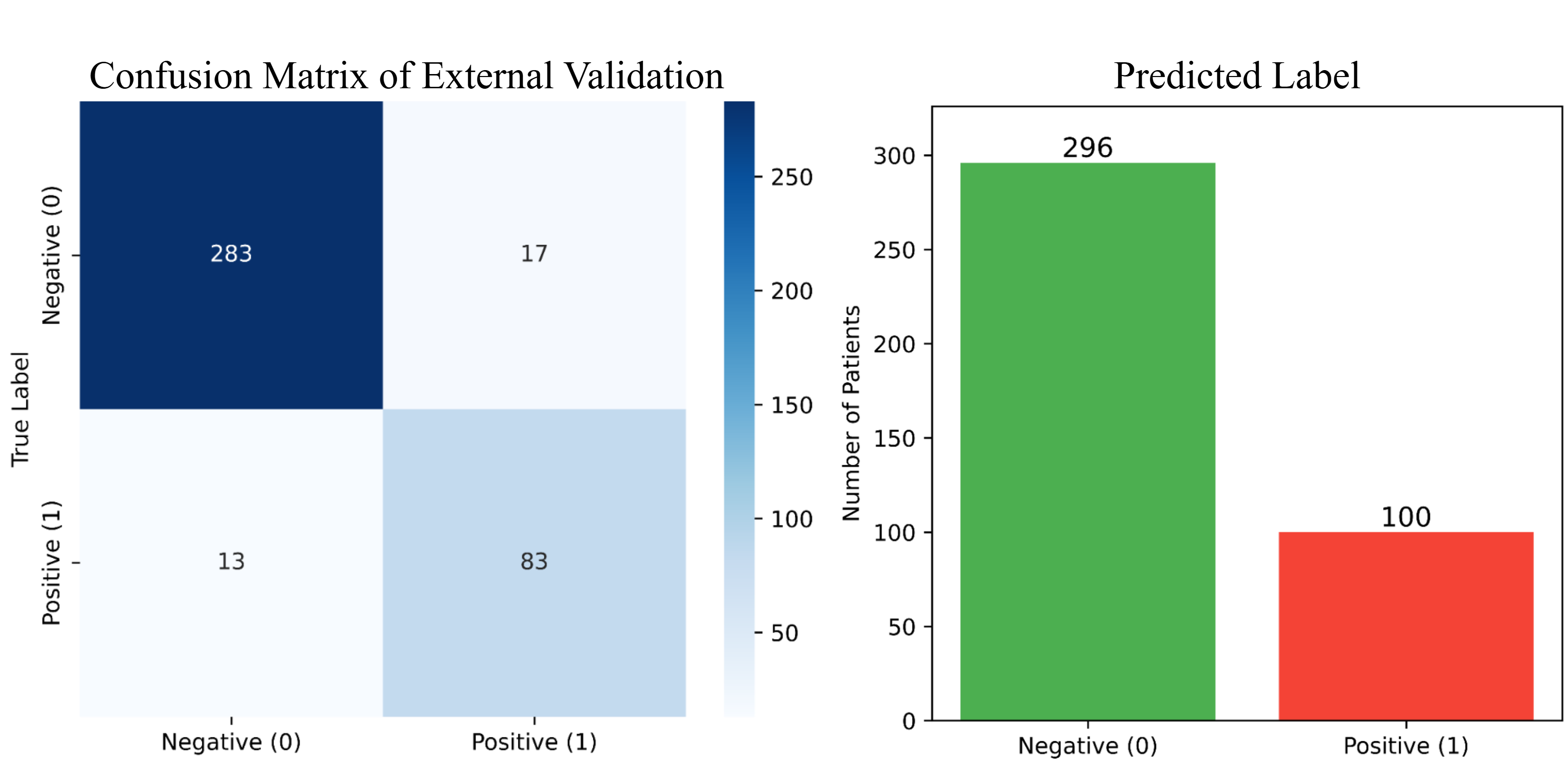}
	\caption{Evaluation results of external validation. Left: confusion matrix of the validation. Right: The histogram of predicted label.}
	\label{fig:external_result}
\end{figure}

\hyperref[fig:external_result]{Figure.\ref{fig:external_result}} demonstrated the evaluation results of our cross-center dataset, with the metrics in \hyperref[tab:external]{Table.\ref{tab:external}}. The results show similar features in our testing set, indicating good generalization capability of our model.

The experimental results presented in Sections 3.1 and 3.2 collectively demonstrate the effectiveness of the proposed CSASN framework in addressing the key challenges of rare thyroid carcinoma recognition. In the comprehensive comparison with seven state-of-the-art baselines (Table \ref{tab:performance_comparison}, Figures \ref{fig:roc_comparison}–\ref{fig:statistical_analysis}), CSASN consistently achieved the highest AUC values across all three subtypes—ATC (0.984), FTC (0.982), and MTC (0.995)—with statistically significant improvements (p < 0.01) and large effect sizes (Cohen’s d: 0.89–1.24). These gains are particularly notable for MTC, where CSASN outperformed the best baseline by 5.6\%, underscoring its capability in capturing subtle discriminative patterns associated with rare malignancies.

The ablation studies further validated the contribution of each architectural component. The cascaded attention module (SE$\rightarrow$CBAM) proved to be the most critical element, as its removal led to the most severe performance degradation across all tasks (e.g., AUC dropped from 0.984 to 0.865 for ATC). This highlights the importance of adaptive feature recalibration in imbalanced and heterogeneous settings. Moreover, the dual-branch design effectively combines local texture details from EfficientNet and global context from ViT, as evidenced by the performance drop when either branch was ablated. The residual multi-scale classifier and dynamic loss weighting further enhanced robustness and generalization, especially under domain shift and class imbalance.

Critically, the model’s strong generalization capability was confirmed through rigorous cross-center validation on a fully independent external dataset. As summarized in Table \ref{tab:external} and Figure \ref{fig:external_result}, CSASN maintained high performance (AUC: 0.9314, Accuracy: 0.9242) when tested on data from Zhejiang Cancer Hospital and Zhongshan Hospital, which were not involved in model development. This external set, comprising 300 benign and 96 FTC samples, presented a distinct domain shift challenge. The model's ability to achieve balanced precision (0.8300) and recall (0.8646) for FTC on this unseen data is a robust testament to the domain-invariant features learned through our MMD-regularized and dynamically weighted optimization strategy. The confusion matrix and prediction histogram in Figure \ref{fig:external_result} further illustrate a well-calibrated and confident classification behavior, mirroring its performance on the internal test set.

In summary, CSASN not only surpasses existing models in diagnostic accuracy on internal data but also demonstrates a more balanced, reliable performance profile and superior generalization to external, heterogeneous clinical environments. These results collectively lay a strong foundation for the following discussion on clinical applicability and limitations.

\section{Discussion}
\label{sec:discussion}

In this study, we proposed the Channel-Spatial Attention Synergy Network (CSASN), a novel deep learning framework tailored for the challenging task of recognizing rare thyroid carcinoma subtypes from ultrasound images. Through extensive experiments on a large, multi-center dataset, we demonstrated that CSASN significantly outperforms a range of strong baseline models, including state-of-the-art CNNs, Vision Transformers, and their hybrids. The comprehensive ablation studies further confirmed the critical contribution of each component within our integrated architecture. Below, we discuss the clinical and technical implications of our findings, situate them within the broader landscape of AI in medical imaging, and acknowledge the limitations of our work.

The superior performance of CSASN, particularly for rare subtypes like MTC (AUC: 0.995), can be attributed to its systematic design that directly addresses the core challenges in thyroid nodule diagnosis.

First, the dual-branch feature extraction module successfully harnesses the complementary strengths of convolutional and transformer architectures. The ablation results clearly show that removing either branch led to a performance decline, but the nature of the decline was subtype-specific. For instance, the removal of the ViT branch (Ablation3) caused a substantial drop in recall for FTC (from 0.9126 to 0.5911), suggesting that the global contextual understanding provided by the Transformer is crucial for identifying the often diffuse and ill-defined borders of FTC \cite{haugen20162015,xing2024multi}. Conversely, for MTC, the removal of the EfficientNet branch (Ablation2) severely impacted the F1-score, underscoring the importance of local, high-resolution texture analysis for characterizing the fine-grained calcifications and homogeneous echotexture often associated with MTC \cite{wells2015revised}. This indicates that our hybrid approach is not merely additive but synergistic, allowing the model to adapt its feature extraction bias based on the distinct morphological characteristics of each subtype.

Second, the cascaded attention refinement module emerged as the most critical component for robust performance, especially under class imbalance. The dramatic performance collapse observed in Ablation1 across all tasks (e.g., AUC dropping from 0.984 to 0.865) highlights that without adaptive feature recalibration, the model struggles to focus on the subtle, discriminative patterns of the rare minority classes. This finding aligns with the clinical diagnostic process, where radiologists sequentially filter information, first identifying diagnostically relevant biomarkers (channel-wise attention) and then localizing their spatial extent (spatial-wise attention) \cite{schlemper2019attention,woo2018cbam}. Our sequential SE$\rightarrow$CBAM design effectively mimics this workflow, amplifying features that are critical for distinguishing rare carcinomas from the predominant benign and PTC cases, thereby mitigating the model's inherent bias toward the majority class.

Third, the dynamic multi-component optimization strategy played a pivotal role in enhancing the model's generalization. By automatically balancing the focal loss (addressing class imbalance), MMD regularization (promoting domain invariance), and BSS term (preventing feature redundancy), the framework ensures stable learning from a complex, multi-center dataset. The use of uncertainty weighting to learn the loss coefficients $\lambda_i$ is particularly advantageous, as it avoids cumbersome manual tuning and allows the model to adapt the learning focus throughout the training process \cite{kendall2018multi}. This is a crucial step toward building models that are robust to the real-world heterogeneity of medical data.

Our work advances the field beyond previous studies that primarily focused on the binary classification of thyroid nodules as benign or malignant \cite{zhu2021thyroid,vadhiraj2021ultrasound,chen2023thyroidnet}. While these studies demonstrated the feasibility of AI in thyroid ultrasound, they often did not address the critical clinical need for subtype stratification, especially for the rare but aggressive carcinomas that carry significantly different prognoses and management plans \cite{haugen20162015,rago2022risk}.

Recent attempts at using more complex architectures have shown promise. For example, hybrid CNN-Transformer models like SimpleHybrid \cite{chen2023thyroidnet} have been explored. However, our CSASN differentiates itself through its deeply integrated and purpose-driven design: (1) Instead of a naive feature concatenation, we employ a cascaded attention mechanism that actively refines the fused features, leading to more discriminative representations. (2) We introduce a residual multi-scale classifier that captures hierarchical patterns, which is more sophisticated than a simple fully-connected layer. (3) Most importantly, we explicitly model and optimize for the triad of challenges—class imbalance, domain shift, and feature heterogeneity—through a dynamically weighted composite loss, a holistic approach not commonly adopted in prior thyroid AI studies \cite{ludwig2023use,yao2024enhancing}.

The significant performance gap between CSASN and all baseline models, including ConViT which incorporates convolutional biases into ViT, validates that our comprehensive, problem-aware architecture delivers tangible improvements over generic or partially optimized solutions.

The high accuracy and robust generalization of CSASN hold significant potential for clinical translation. An AI system capable of reliably flagging rare thyroid cancer subtypes could serve as a valuable decision-support tool for sonographers and endocrinologists. It may help reduce the rate of missed diagnoses of rare carcinomas, which is a serious concern given their low prevalence and often atypical appearance \cite{ludwig2023use,durante2018diagnosis}. By highlighting suspicious nodules that warrant more aggressive diagnostic workup (e.g., FNA or specific molecular testing), such a system could contribute to earlier detection and more personalized treatment planning for patients with FTC, MTC, or ATC.

Furthermore, the model's demonstrated resilience to cross-center domain shift is a critical step toward real-world deployment. The ability to maintain performance on an independent external test set, despite differences in ultrasound devices and imaging protocols, suggests that CSASN learns biologically relevant features rather than institution-specific artifacts. This aligns with the growing emphasis in medical AI on developing models that generalize across diverse healthcare settings \cite{zhang2020generalizing,zhou2022domain}.

Despite the promising results, our study has several limitations. First, the external test set, while valuable for assessing generalization, did not contain samples for MTC and ATC. This limits our ability to fully validate the model's performance on these rarest subtypes in an unseen clinical environment. Future work must include external testing with a complete spectrum of subtypes.

Second, the current framework operates on a single ultrasound image. In clinical practice, radiologists make diagnoses based on dynamic sweeps through the nodule and incorporate rich clinical context (e.g., patient age, sex, serum calcitonin levels for MTC). Integrating video data and multimodal clinical information represents a important and natural extension of this work.

Third, while our ablation studies are comprehensive, the computational cost of the dual-branch architecture is non-negligible. For future real-time applications, exploring model distillation or neural architecture search to optimize the efficiency of CSASN would be highly beneficial.

Finally, as with all AI diagnostic tools, prospective clinical trials are the ultimate step to validate the efficacy and utility of CSASN in real-world clinical workflows, assessing its impact on diagnostic confidence, workflow efficiency, and, ultimately, patient outcomes.

\section{Conclusion}
\label{sec:conclusion}

In conclusion, we have presented CSASN, a robust deep-learning framework for the recognition of rare thyroid carcinoma subtypes from ultrasound images. By synergistically combining a dual-branch CNN-Transformer backbone, a cascaded channel-spatial attention module, a residual multi-scale classifier, and a dynamic multi-task loss function, CSASN effectively addresses the intertwined challenges of class imbalance, morphological heterogeneity, and cross-center domain shift. Our extensive evaluations on a multi-institutional dataset confirm that the proposed model achieves state-of-the-art performance and exhibits strong generalization ability. This work provides a solid and practical pathway toward AI-assisted precision diagnosis in thyroid oncology, with the potential to improve the detection and management of clinically significant, rare thyroid cancers.

\section*{Acknowledgement}
Thanks to Mr. Jiayuan She from University of California, Santa Cruz, Zihan Liang and Shukun Geng from Xi'an Jiaotong-Liverpool University for their support during this research.

\section*{Ethics Approval}
This study is approved by the Ethical Committee of Shanghai Tenth People's Hospital (Approval No. 22XJS36) with the informed consent of all participants.

\section*{Authors Contribution Statement}
\textbf{Peiqi Li}: Conceptualization, Software, Methodology, Visualization, Writing-draft, review \& editing, Project Administration; \textbf{Yincheng Gao, Haojie Yang}: Data Curation, Methodology; \textbf{Renxing Li}: Project Administration; \textbf{Yunyun Liu, Boji Liu, Jiahui Ni, Ying Zhang, Yulu Wu}: Data Curation and Management, \textbf{Xiaowei Fang}: Writing-draft; \textbf{Jiangang Chen, Liping Sun, Lehang Guo}: Supervision, Validation, Writing-review.

\section*{Declaration of Competing Interest}
The authors declare that they have no known competing financial interests or personal relationships that could have appeared to influence the work reported in this paper.

\section*{Code and Data Availability}
The datasets used during the current study are not publicly available due to ethical or privacy concerns. The code generated for this study is available from the corresponding author upon reasonable request.

\bibliographystyle{elsarticle-harv}
\bibliography{Reference.bib}
\end{document}